# CONTRACT SIGNATURE USING QUANTUM INFORMATION

Paulo Benício Melo de Sousa and Rubens Viana Ramos
benicio@deti.ufc.br        rubens@deti.ufc.br
*Departamento de Engenharia de Teleinformática – Universidade Federal do Ceará - DETI/UFC*
*C.P. 6007 – Campus do Pici - 60755-640 Fortaleza-Ce Brasil*

This paper describes how to perform contract signature in a fair way using quantum information. The protocol proposed permits two partners, users of a communication network, to perform a contract signature based on the RSA security. The authentication of the signers is based on the use of a non-local XOR function of two classical bits.

## 1. Introduction

Contract signature is an important part of security information area with many practical applications. The main goal is to provide a mechanism in which two participants can exchange a sequence of bits, representing their signatures, in a fair way, that is, one of them will send his/her signature only if he/she is sure that will also receive the partner' signature. Hence, a good contract signature protocol must guarantee fairness for both signers.

In this paper, we propose a model for contract signature using quantum information. The protocol proposed is based on the non-local XOR function between two classical bits. This work is outlined as follows: Section 2 describes how to implement the non-local XOR function between two classical bits; Section 3 shows the protocols for quantum contract signature; at last, conclusions are presented in Section 4.

## 2. Non-local XOR function between two classical bits

The non-local XOR function between two classical bits was proposed in [1] and here we briefly explain it. Firstly, we consider that there are three authorized parties of the communication, Alice, Bob and Charlie, sharing the following maximally entangled tripartite of qubit state (obtained by application of a Hadamard gate in each individual state of a tripartite GHZ state):

$$|\psi\rangle = \frac{1}{2}\left(|000\rangle_{ABC} + |011\rangle_{ABC} + |110\rangle_{ABC} + |101\rangle_{ABC}\right) \qquad (1)$$

Considering $\rho_A$, $\rho_B$ and $\rho_C$ as the individual parts of the total state $|\psi\rangle$, the non-local XOR function between two classical bits, represented by $K$ (belonging to Alice) and $R$ (belonging to Bob) can be achieved using the quantum circuit shown in Fig. 1.

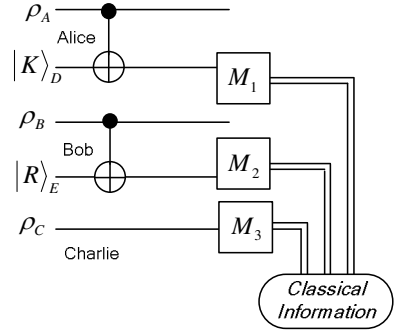

Fig. 1 - Quantum circuit for non-local XOR function between two classical bits. $M_{1-3}$ are measurers.

In Fig. 1, $M_1$, $M_2$ and $M_3$ are measurers. The initial and final states are respectively given by:

$$|\Psi_{in}\rangle = |KR\rangle_{DE} \otimes \frac{1}{2}\left(|000\rangle_{ABC} + |011\rangle_{ABC} + |110\rangle_{ABC} + |101\rangle_{ABC}\right) \qquad (2)$$

$$|\Psi_{out}\rangle = \frac{1}{2}\left\{\begin{array}{l}|00\rangle_{AB}|KR\rangle_{DE}|0\rangle_C + |01\rangle_{AB}|K\overline{R}\rangle_{DE}|1\rangle_C + \\ |11\rangle_{AB}|\overline{KR}\rangle_{DE}|0\rangle_C + |10\rangle_{AB}|\overline{K}R\rangle_{DE}|1\rangle_C\end{array}\right\} \qquad (3)$$

When the qubits $D$, $E$ and $C$ are measured, by Alice, Bob and Charlie, respectively, the values {110, 101, 000, 011}$_{DEC}$ are obtained only if qubits $K$ and $R$ are equal. On the other hand, if $K$ and $R$ are not equal only the values {100, 111, 010, 001}$_{DEC}$ can be obtained by the measurement. Hence, the non-local XOR function can be described as follows:

- Alice measures qubit $D$ and she sends her result to Charlie using one classical bit.
- Bob measures qubit $E$ and he sends his result to Charlie using another classical bit.
- Charlie, by its turn, performs a measurement in his qubit. Knowing those three classical information, Charlie can know if $K$ and $R$ are equal or not, that is $K \oplus R$.

The bits sent by Alice and Bob inform to Charlie not the values of $K$ and $R$, but if $K$ and $R$ are equal or not to the individual states $\rho_A$ and $\rho_B$, respectively.

## 3. Contract signature protocol

The proposed contract signature protocol is strongly based on the non-local XOR function discussed in Section 2. Alice and Bob are the contractors and Charlie is the arbitrator that works as a TTP (trustable third part). This problem has some particularities:

- *Proper signature*: Alice and Bob's signatures are related to the contract, that is, a signature is a function of the contract. This means that the usual idea of accessing a database in order to get a signature (using, for instance, a *Key Distribution Center* – KDC) is not considered in this work.
- *Fairness*: If Alice and Bob have intention of sign a contract, they have to use their private functions over the common message (contract) and exchange the information simultaneously. Therefore, there is no risk that one of them sending his/her signature without receiving the partner's signature.
- *Validation:* Both candidates can evaluate if the signature received is correct in order to validate the protocol;
- *Abuse-freeness*: Charlie does not have access to the Alice's and Bob's signatures using classical computation.

Basically, the proposed protocol has three stages:

- *The initialization*: when Alice and Bob choose their signatures;
- *Simultaneous message passing*: when Charlie receives the information of both parties that will be passed to the respective parties using the non-local XOR function;
- *The validation*: when, finally, Alice and Bob obtain the counterpart signature in a non-secure scenario (for instance, if neither Alice nor Bob trust each other), both can evaluate the information received and to send to Charlie the confirmation, or not, of the validity of the data.

In the protocol proposed, we consider that Alice and Bob use the RSA algorithm [2] in order to create their signatures. The security of RSA is based on the intractability of integer factorization using classical computers. Briefly, two large prime numbers $p$ and $q$ are chosen generating $n = p \times q$ and $z = (p - 1) \times (q - 1)$. Then, a prime number $d$ relative to $z$ is chosen and it is calculated a number $e$ such that $e \times d = 1 \bmod z$. Finally, the signature $s$ of a contract $m$ is calculated as $s = m^d \bmod n$ and, reversely, in order to check if the signature is correct, one calculates $m = s^e \bmod n$. Considering this algorithm, the proposed protocol for contract signature is described as follows:

*Initialization stage:*

1. First of all, Alice (*A*) and Bob (*B*) agree with a specific contract that is represented by a message *m*. This contract is not required to be secret and, for our purposes, it is managed by Charlie (playing the role of a TTP).

2. As required by RSA, *A* and *B* choose, respectively, the values of $\{n_A, e_A, d_A\}$ and $\{n_B, e_B, d_B\}$. These values are used by the RSA scheme as explained before: $s_k = m^{d_k} \bmod n_k$ and $m = s_k^{e_k} \bmod n_k$, $k = \{A, B\}$, where $s_{A(B)}$ is the signature of *A(B)*.

*Simultaneous message passing stage:*

3. Using the scheme of Section 2, it is possible to implement a secure quantum protocol that permits Charlie to know the XOR function between the signatures of Alice and Bob (concurrence), $s_{AB} = s_A \oplus s_B$.

4. Charlie sends $s_{AB}$ to Alice and Bob. Since Charlie is not one of the interested parties, he can be considered neutral and that he will, in fact, to send the same information to both signers.

*The validation stage:*

5. Now, having $s_{AB}$, *A* and *B* can get the value of the counterpart signature performing a XOR operation between $s_{AB}$ and their own signatures. In this way, Alice knows which bits she has to flip in her signature in order to transform it in Bob's one. Similarly, Bob obtains Alice' signature.

6. In order to check if Alice (Bob) has sent a valid signature, Bob (Alice) uses $s_A$ ($s_B$) and the public information $n_A$ and $e_A$ ($n_B$ and $e_B$) in order to recover the original contract *m*. If the comparisons are positives, they inform Charlie that the contract signature can be validated. If not, the one who is claiming having received a false signature aware Charlie about it.

In order to check the security let us consider two situations: 1) Alice claims that Bob has sent a false signature and Bob has, in fact, sent a false signature. 2) Alice claims that Bob has sent a false signature but Bob sent the correct one. The actions of Charlie for both cases are:

1. Charlie asks Alice to send to him Bob's and her signatures, $S_{BA}$ and $S_{AA}$ ($S_{xy}$ means $s_x$ according to $y$). Further, Charlie also requires $d_A$.

2. Charlie checks if: I) $m = s_{BA}^{e_B} \mod n_B$; II) $S_{BA} \oplus S_{AA} = S_{AB}$; III) $s_{AA} = m^{d_A} \mod n_A$; IV) $m = s_{AA}^{e_A} \mod n_A$. If I is false and II-IV are positive, then Charlie believes that Bob sent a false signature. Otherwise, Charlie will believe that Alice is lying.

The procedure is analogous if Bob (lying or not) claims that Alice has sent a false signature.

## 4. Conclusions

In this work, it was presented a scheme for contract signature using quantum information. Basically, the signatures are generated according the contract to be considered and the simultaneous exchange of information is done using the non-local XOR function between two classical bits. The protocol is also based in the RSA algorithm and considers that Charlie is a TTP responsible to send the XOR information between the parties. At last, it was presented a security analysis that permits Charlie to verify if one of the parties sent a false signature or lied saying the partner have done it. In the first case the contract is cancelled while in the last case it remains valid. Moreover, penalties can be applied to the user that was caught cheating.

**Acknowledgements**

This work was supported by the Brazilian agency FUNCAP.